\documentclass{PoS}
\usepackage{bbm}
\newcommand{\Tr}{{\rm Tr}}

\newcommand{\PM}{\mathbbm{P}_M}

\let\OLDthebibliography\thebibliography
\renewcommand\thebibliography[1]{
  \OLDthebibliography{#1}
  \setlength{\parskip}{0pt}
  \setlength{\itemsep}{0pt plus 0.0ex}
}
\title{{\vspace{-15mm} \normalsize\hfill{\small DESY 14-207}}\\[15mm]
{\vspace{-18mm}\normalsize\hfill{\small HU-EP-14/46}}\\[12mm] 
{\vspace{-15mm}\normalsize\hfill{\small SFB/CPP-14-85}}\\[10mm] Comparison of different lattice
definitions of the topological charge}

\ShortTitle{Lattice definitions of the topological charge}

\author{\speaker{Krzysztof Cichy}$^{abc}$\thanks{We acknowledge useful discussions with Karl Jansen at
various stages of this project.}, Arthur Dromard$^{b}$, Elena
Garcia-Ramos$^{ad}$, Konstantin Ottnad$^{ef}$, Carsten Urbach$^f$, Marc Wagner$^b$, Urs Wenger$^g$, Falk
Zimmermann$^f$\\ 
\llap{$^a$}John von Neumann Institute for Computing (NIC), DESY, Platanenallee 6, 15738 Zeuthen,
Germany\\
\llap{$^b$}Goethe-Universit\"at Frankfurt am Main, Institut f\"ur Theoretische
Physik, Max-von-Laue-Stra{\ss}e 1, D-60438 Frankfurt am Main, Germany
\thanks{A.D.\ and M.W.\ acknowledge
support by the Emmy Noether Programme of
the DFG (German Research Foundation), grant WA 3000/1-1. This work was
supported in part by the Helmholtz International Center for FAIR within
the framework of the LOEWE program launched by the State of Hesse.}\\
\llap{$^c$}Faculty of Physics, Adam Mickiewicz University, Umultowska
85, 61-614 Pozna\'{n}, Poland\\
\llap{$^d$}Humboldt Universit\"at zu Berlin,  Newtonstrasse 15, 12489 Berlin, Germany\\
\llap{$^e$}Department of Physics, University of Cyprus, P.O. Box 20537, 1678 Nicosia, Cyprus\\
\llap{$^f$}Helmholtz-Institut f\"ur Strahlen- und Kernphysik (Theorie) 
and Bethe Center for Theoretical Physics, Universit\"at Bonn, 53115 Bonn, Germany\\
\llap{$^g$}Albert Einstein Center for Fundamental Physics, Institute for
Theoretical Physics, University of Bern, Sidlerstrasse 5, 3012 Bern, Switzerland\\
E-mail: \email{krzysztof.cichy@desy.de}, \email{dromard@th.physik.uni-frankfurt.de},
\email{elena.garcia.ramos@desy.de}, \email{karl.jansen@desy.de},
\email{ottnad@hiskp.uni-bonn.de}, \email{urbach@hiskp.uni-bonn.de},
\email{mwagner@th.physik.uni-frankfurt.de}, \email{wenger@itp.unibe.ch},
\email{fzimmermann@hiskp.uni-bonn.de}
}

\abstract{
We present a comparison of different definitions of the topological charge on the lattice, using a
small-volume ensemble with 2 flavours of dynamical twisted mass fermions. The investigated definitions
are: index of the overlap Dirac operator, spectral projectors, spectral flow of the Hermitian
Wilson-Dirac operator and field theoretic with different kinds of smoothing of gauge fields (HYP and APE
smearings, gradient flow, cooling). We also show some results on the topological susceptibility.
\begin{center}
\vspace*{-0.5cm}
\includegraphics
[width=0.15\textwidth,angle=0]
{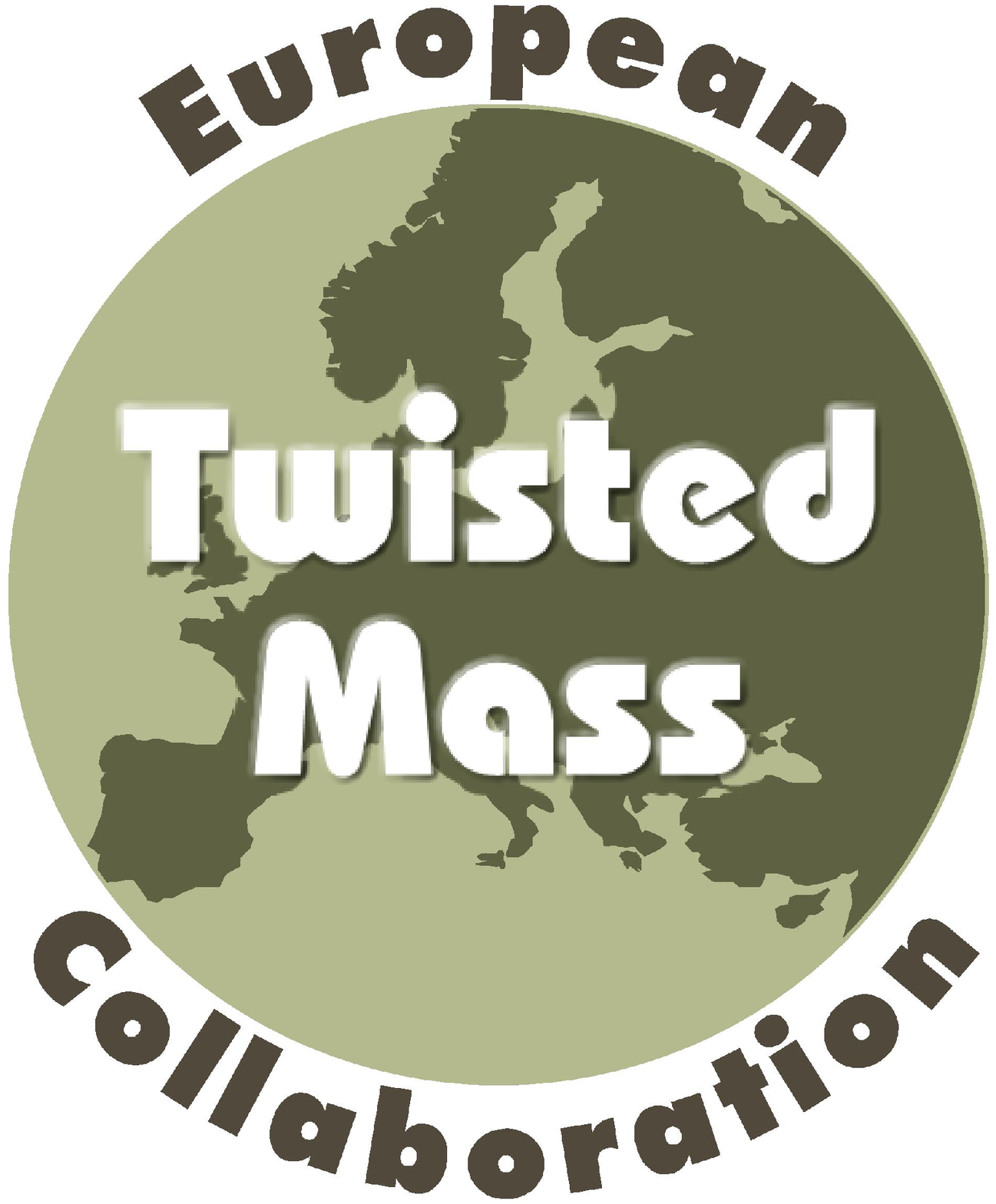}
\end{center}
\vspace*{-0.2cm}
}

\FullConference{The 32nd International Symposium on Lattice Field Theory,\\
		23-28 June, 2014\\
		Columbia University New York, NY}

\begin{document}

\section{Introduction}
\vspace*{-0.3cm}
Gauge fields in QCD can be characterized by an integer number, the
topological charge, that is related to their topological properties. Such properties play a central role
in understanding the structure of the QCD vacuum. At the same time, topological excitations do not
appear in perturbation theory -- hence, a non-perturbative approach is essential to understand topology
in QCD. The most successful non-perturbative approach is to use the lattice as a regulator.
However, for many years it was unclear how to properly define the topological charge on the
lattice. The proposed definitions had their flaws or were impractically expensive numerically.
Only in the recent years, significant progress has been achieved with the advent of new theoretically
sound definitions.

The aim of this proceeding and an upcoming paper \cite{Cichy2014} is to critically compare the various
definitions of the topological charge and argue that the theoretical progress of the past few years
finally makes it possible to resolve the topological issues in QCD in an unambiguous way.
The outline of the current proceeding is as follows. In Sec.~2, we shortly review the definitions that
we used. Sec.~3 shows our comparison of different definitions and discusses it. Sec.~4
concludes.

\begin{table}[t!]
  \centering
  \begin{tabular}[]{ccccc}
    nr & full name & smearing & short name & type\\
\hline
1 & index of overlap Dirac operator $s=0.4$ & -- & index nonSmear $s=0.4$ & F\\
2 & index of overlap Dirac operator $s=0.0$ & -- & index nonSmear $s=0$ & F\\
3 & index of overlap Dirac operator $s=0.0$ & HYP1 & index HYP1 $s=0$ & F\\
4 & Wilson-Dirac op. spectral flow $s=0.75$ & HYP1 & SF HYP1 $s=0.75$ & F\\  
5 & Wilson-Dirac op. spectral flow $s=0.0$ & HYP1 & SF HYP1 $s=0.0$ & F\\ 
6 & Wilson-Dirac op. spectral flow $s=0.5$ & HYP5 & SF HYP5 $s=0.5$ & F\\
7 & Wilson-Dirac op. spectral flow $s=0.0$ & HYP5 & SF HYP5 $s=0.0$ & F\\  
8 & spectral projectors $M^2=0.00003555$ & -- & spec. proj. $M^2=0.0000355$ & F\\
9 & spectral projectors $M^2=0.0004$ & -- & spec. proj. $M^2=0.0004$ & F\\
10 & spectral projectors $M^2=0.0010$ & -- & spec. proj. $M^2=0.0010$ & F\\
11 & spectral projectors $M^2=0.0015$ & -- & spec. proj. $M^2=0.0015$ & F\\
12 & improved field theoretic & GF$t_0$ & GF flow time $t_0$ & G\\
13 & improved field theoretic & GF$2t_0$ & GF flow time $2t_0$ & G\\
14 & improved field theoretic & GF$3t_0$ & GF flow time $3t_0$ & G\\
15 & improved field theoretic & -- & impr. FT nonSmear & G\\
16 & improved field theoretic & HYP10 & impr. FT HYP10 & G\\
17 & improved field theoretic & HYP40 & impr. FT HYP40 & G\\
18 & improved field theoretic & APE10 & impr. FT APE10 & G\\
19 & improved field theoretic & APE30 & impr. FT APE30 & G\\
20 & naive field theoretic & APE10 & naive FT APE10 & G\\
21 & naive field theoretic & APE30 & naive FT APE30 & G\\
22 & improved field theoretic & impr. cool. & impr. FT impr. cool. 10 & G\\
23 & improved field theoretic & impr. cool. & impr. FT impr. cool. 30 & G\\
24 & naive field theoretic & impr. cool. & naive FT impr. cool. 10 & G\\
25 & naive field theoretic & impr. cool. & naive FT impr. cool. 30 & G\\
26 & improved field theoretic & basic cool. & impr. FT basic cool. 10 & G\\
27 & improved field theoretic & basic cool. & impr. FT basic cool. 30 & G\\
28 & naive field theoretic & basic cool. & naive FT basic cool. 10 & G\\
29 & naive field theoretic & basic cool. & naive FT basic cool. 30 & G\\
\end{tabular}
 \caption{\label{tab:defs} The relevant characteristics of each topological charge definition. For each
definition, we give a number, full name, type of smearing of gauge fields (-- = no smearing, HYP$n$ = $n$
iterations of HYP smearing, APE$n$ = $n$ iterations of APE smearing, GF$t$ = gradient flow at flow time
$t$, cool. = improved or basic cooling, explained in text), short name (used in plots) and definition
type (G=gluonic,
F=fermionic).}
\end{table}

\section{Short review of lattice definitions of the topological charge}
\vspace*{-0.3cm}
The lattice definitions of the topological charge (denoted by $Q$) that we have used are summarized in
Tab.~\ref{tab:defs}. Below, we include their short review. For a more comprehensive
discussion, we refer to our upcoming paper \cite{Cichy2014} and to original papers.
\begin{itemize}
\item \textbf{Index of the overlap Dirac operator.} Chirally symmetric fermionic discretizations
allow exact zero modes of the Dirac operator. The famous Atiyah-Singer index theorem
relates the topological charge to the number of zero modes of the Dirac
operator: $Q = n_{-} - n_{+}$, where $n_\pm$ are, respectively, the number of zero modes in the positive
and in the negative chirality sector. This definition is theoretically sound \cite{Hasenfratz:1998ri}, it
does not require renormalization and it provides integer values of the topological charge. It is also
unique, up to the dependence on the $s$ parameter of the kernel of the overlap Dirac operator (which is,
however, only a cut-off effect). This definition has been known for many years and its only disadvantage
is practical -- the cost of using overlap fermions is approximately two orders of magnitude larger than
for e.g. Wilson fermions.
\item\textbf{Wilson-Dirac operator spectral flow}. This definition is equivalent to the index
of the overlap Dirac operator \cite{Itoh:1987iy}. One considers the mass dependence of the eigenvalues of
the Hermitian Wilson-Dirac operator $D^\dagger D+m^2$. Tracing the evolution of each eigenvalue, one
counts the number of net crossings of zero in a given mass range, i.e. the difference of crossing from
above and from below. This net number of crossings corresponds to the index of the overlap operator at a
corresponding value of the $s$ parameter. As such, this definition has all the advantages of the overlap
index definition, at a lower cost. However, in practice it might be difficult to resolve the crossings in
an unambiguous way (in particular at coarse lattice spacings) and hence additional computations may be
needed to clarify these ambiguities.
\item\textbf{Spectral projectors.}
This is another fermionic definition, introduced in Refs.~\cite{Giusti:2008vb,Luscher:2010ik}. It
defines a projector to the subspace of eigenmodes of $D^\dagger D$ with eigenvalues below a certain
threshold $M^2$. Using this projector, one can stochastically evaluate the topological
charge $Q=\Tr\,\{\gamma_5\PM\}$. For chirally symmetric fermions, such a definition is again equivalent
to the index (i.e.\ it is a stochastic way of counting the zero modes), while for non-chirally
symmetric fermions it still gives a clean definition, although
chirality of modes is $\pm1+\mathcal{O}(a^2)$ and renormalization with $Z_S/Z_P$ is needed.
In the spectral projector formulation, the topological charge depends on the $M$ parameter, however,
this dependence is a cut-off effect.
Due to the stochastic ingredient and to cut-off effects, the extracted value of the topological charge
is non-integer, which, however, poses no theoretical problem.
For the computation of the topological susceptibility using this approach, see Ref.~\cite{Cichy:2013rra}.
\item\textbf{Field theoretic.}
This definition is conceptually very different from the fermionic ones discussed above, as it is
purely gluonic. Historically, it is the oldest definition, since it is usually cheap to compute and it is
very natural, since in the continuum, the topological charge is given by $Q =
\frac{1}{32\pi^2}\int {\rm d}^4x\,\epsilon_{\mu\nu\rho\sigma} {\rm
tr}[F_{\mu\nu}(x)F_{\rho\sigma}(x)]$.
On the lattice, one has to choose some discretization for the field-strength tensor $F_{\mu\nu}$ (we
apply the simplest discretization using only plaquettes (``naive''), as well as an
improved version using also $2\times2$ and $3\times3$ Wilson loops).
This leads to short-distance singularities that need to be removed using some smoothing (``filtering'')
procedure. The commonly used methods to do this are shortly discussed below.
\begin{itemize}
\vspace*{-0.2cm}
\item \textbf{Gradient flow} -- the method recently introduced by L\"uscher
\cite{Luscher:2010iy} is a rigorous way of smoothing of gauge fields and it can be shown to be free of
divergences to all orders in perturbation theory \cite{Luscher:2011bx}. As such, it provides a
theoretically sound definition of the topological charge, which requires no renormalization. It is also
much cheaper with respect to the index of the overlap operator. We consider the gradient flow using the
Wilson gauge action at flow times $t_0$, $2t_0$ and $3t_0$.
\vspace*{-0.1cm}
\item \textbf{Cooling} -- an iterative minimization of the lattice action, eliminates
rough topological
fluctuations, but keeps large instantons unchanged and decreases the UV noise
\cite{Berg:1981nw,Iwasaki:1983bv,Teper:1985rb,Ilgenfritz:1985dz}. This procedure has been extensively
used in the past. It can be thought of as a discrete version of gradient flow. It can be matched to
gradient flow \cite{Bonati:2014tqa}, which gives it more theoretical justfication. No renormalization is
required.
We consider 10 or 30 steps of 2 versions of cooling: 
\begin{itemize}
\item basic cooling -- lattice action to minimize contains only $1\times1$ plaquettes,
\item improved cooling -- also $2\times2$ and $3\times3$ Wilson
loops \cite{de Forcrand:1997sq}.
\end{itemize}
\vspace*{-0.1cm}
\item \textbf{APE/HYP smearing} -- a discrete procedure that
eliminates short-distance fluctuations, introduced in Refs.~\cite{Albanese:1987ds,Hasenfratz:2001hp}. It
requires additive and multiplicative renormalization. For HYP, we used $\alpha_1 = 0.75$, $\alpha_2 =
0.6$ and $\alpha_3=0.3$. For APE, $\alpha=0.45$. Again, we consider 10 or 30 steps of smearing.
\end{itemize}
\end{itemize}

\vspace*{-0.3cm}
\section{Results}
\vspace*{-0.1cm}
\subsection{Lattice setup}
We use a single ensemble of maximally twisted mass fermions \cite{Frezzotti:2000nk,Frezzotti:2003ni}
with $N_f=2$ flavours, with $\beta=3.9$, $L/a=16$, $a\mu=0.004$ (corresponding to a pion mass of approx.
340 MeV in infinite volume), $a\approx0.079$ fm, hence a small physical volume of $L\approx1.3$ fm. For
more details about this ensemble, we refer to Ref.~\cite{Cichy:2010ta}. In our
upcoming paper \cite{Cichy2014}, we will also consider other ensembles to study the behaviour of
topological charge and topological susceptibility towards the continuum limit.

\begin{figure}[t!]
\hspace*{-2cm}
\includegraphics[width=0.82\textwidth,angle=270]{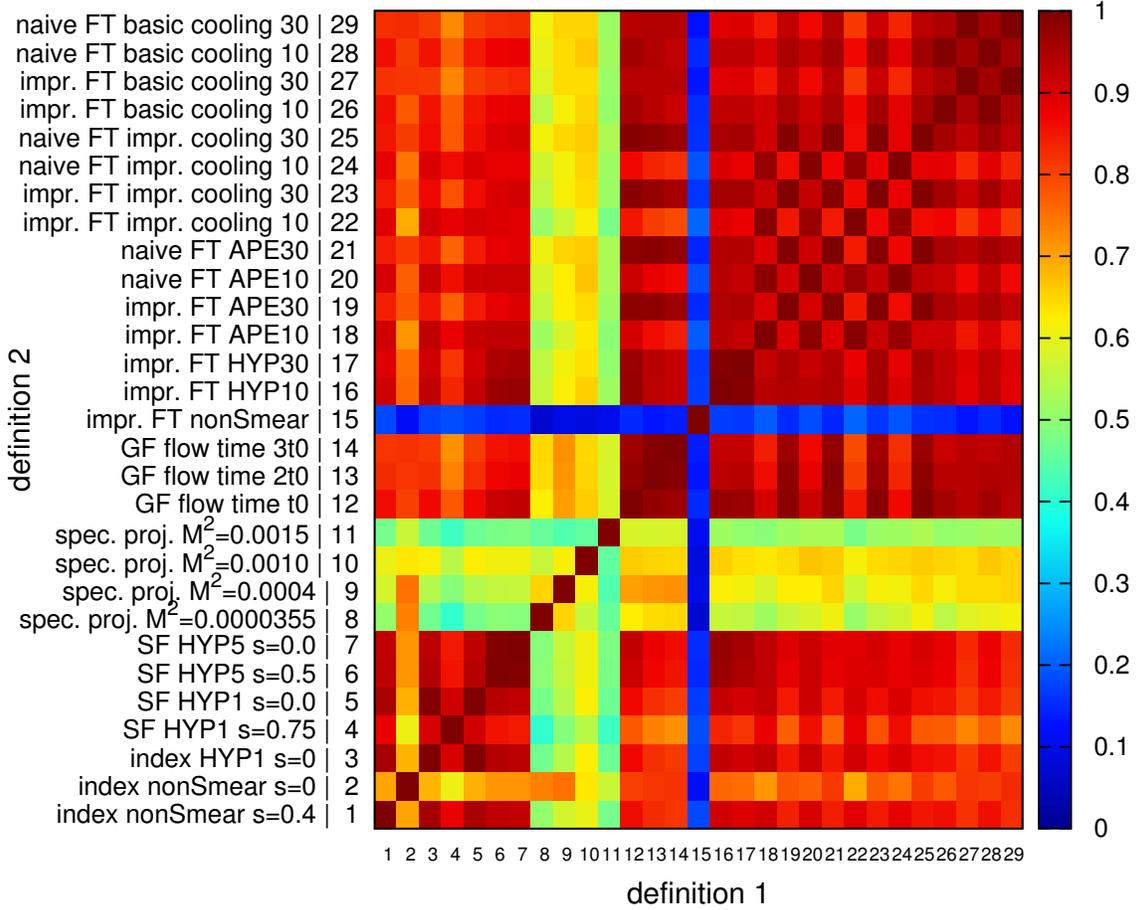}
\caption{\label{fig:b40-corr} Correlation matrix between different definitions of the topological
charge. Cool colours correspond to low correlations (blue: approx. no correlation), while warm colours
denote large correlations (dark red: approx. perfect correlation). The employed definitions are all the
ones contained in Tab.~\protect\ref{tab:defs}.}
\end{figure}

\vspace*{-0.2cm}
\subsection{Correlation between different definitions}
Our main result is the correlation matrix between different definitions of the topological charge,
presented as a colour-coded graph in Fig.~\ref{fig:b40-corr}. We summarize here some conclusions.
\begin{itemize}
\item The general level of correlations between different definitions is very high -- between 80\% and
nearly 100\%. In particular, the topological charge extracted with the field theoretic definition but
with different kinds of smoothing is typically 90-100\% correlated.
\item Exceptions to these rule are: field theoretic definition without any smoothing (where basically
noise is extracted) and spectral projectors, for which the results are contaminated by the stochastic
ingredient (correlation of 40-70\% with other definitions).
\item There is rather significant dependence of the index definition (extracted both from overlap and
Wilson-Dirac spectral flow) on the mass parameter $s$. In certain cases, the correlation can drop even
below 60\%. This can be attributed to very bad locality properties of the overlap operator for some
values of $s$, in particular $s=0$ (nonSmear) yields a very small decay rate of the overlap operator --
for more details, see Ref.~\cite{Cichy:2012vg}.
\item The spectral projector definition shows a significant dependence on the parameter $M^2$. For the
lowest considered $M^2$, the mode number is around 5, which means that not all zero modes are counted.
For higher values of $M^2$ the correlation with other definitions increases up to some value where noise
starts to dominate and the correlation again decreases.
\item Gradient flow shows very similar results at different flow times ($t_0$, $2t_0$ and $3t_0$).
Comparing to other definitions, better correlations are always observed with 30 rather than 10 steps of
smearing or improved cooling, for both the naive and improved definition of the topological charge. This
is not true for basic cooling, where correlations tend to be very similar for 10 and 30 cooling steps.
This results from the underlying dependencies between gradient flow and cooling/smearing. We will
comment more on this issue in the concluding part of this proceeding. 
\end{itemize}

\subsection{Topological susceptibility from different definitions}
\begin{figure}[t!]
\begin{center}
\includegraphics[width=0.715\textwidth,angle=270]{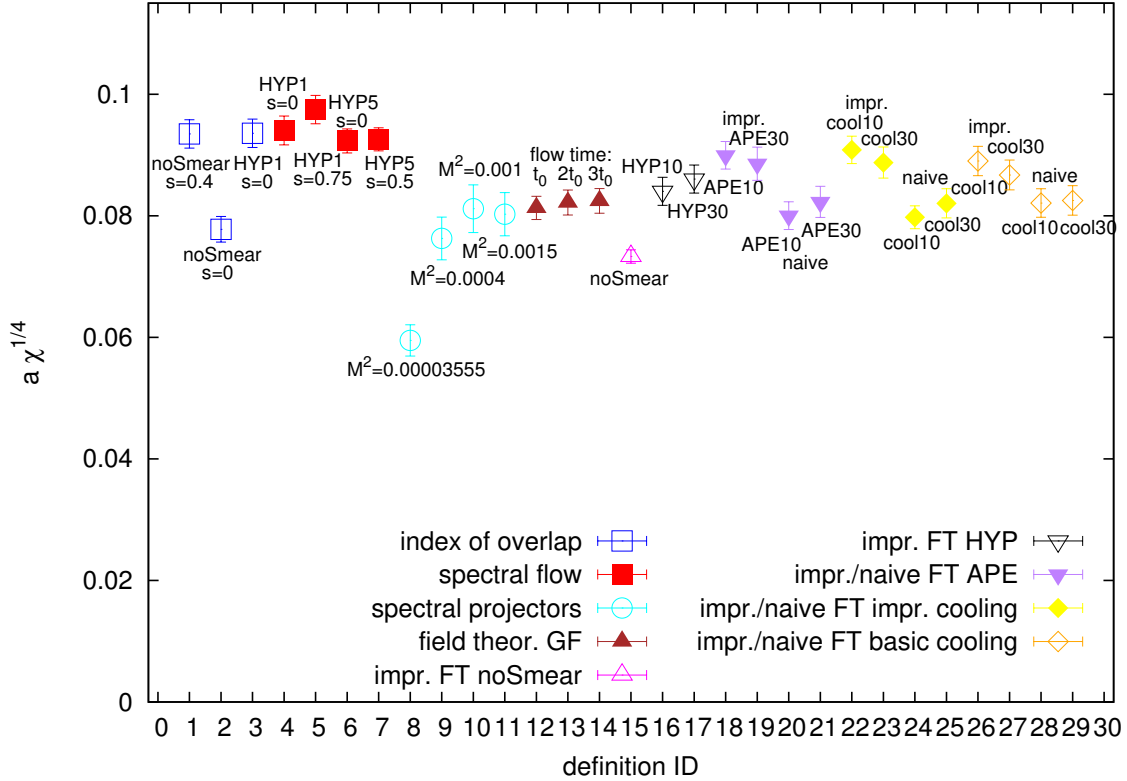}
\caption{\label{fig:b40-chi} Topological susceptibility from different definitions. All the values are
renormalized, except for HYP/APE, where the renormalization was not yet computed systematically, but the
renormalization factor was estimated to be rather small, corresponding to a less than 5-percent
shift
upwards of $a\chi^{1/4}$. The employed definitions are all the ones contained in
Tab.~\protect\ref{tab:defs}.}
\end{center}
\end{figure}

In Fig.~\ref{fig:b40-chi}, we show the results for the topological susceptibility for our test ensemble.
The values of the susceptibility are for almost all methods within approx. 10\% of the value
\mbox{$a\chi^{1/4}\approx0.09$}. Hence, although the values are not strictly compatible for different
definitions, the differences can plausibly be attributed to cut-off effects.
The values which are more than 10\% off from 0.09 result from some flaws of the employed definitions:
bad locality of the overlap Dirac operator (index nonSmear $s=0$), too small value of $M^2$ in spectral
projectors or large UV fluctuations for the field theoretic definition on non-smoothed gauge fields.

\section{Conclusion}
We presented preliminary results of our comparison of different definitions of the topological
charge. We found that for all reasonable definitions (without easily identifiable flaws), the correlation
between the values of the topological charge on our test ensemble is rather high and the values of the
topological susceptibility are in a relatively small range.

In the near future, we plan to extend this investigation to finer lattice spacings to observe the
presumed increase of correlation towards the continuum limit.
Here, we would like to draw attention to an important aspect related to taking the continuum limit.
This is rather unambiguous in the case of the fermionic definitions.
However, for the field theoretic definition, one encounters the problem of matching the different
lattice spacings.
This can be done fully systematically if the smoothing procedure is gradient flow -- one can
consider the topological charge at a fixed flow time, e.g. $t_0$.
A problem appears in the case of cooling/smearing, which can not be considered as rigorous procedures in
the quantum field theoretic sense, since they are discrete.
Nonetheless, to overcome this problem, one can perform the matching of cooling/smearing to gradient
flow, thus defining the correspondence between flow time and the number of cooling/smearing steps.
This has been done for the simplest case of gradient flow with the Wilson plaquette action and basic
cooling \cite{Bonati:2014tqa}, but it can be extended to more general cases.
Thus, to investigate the increase of correlation towards the continuum limit, we plan to use a strategy
to relate the number of cooling/smearing steps such that they correspond to the same flow time at
different lattice spacings.
We emphasize that only in this way the approach to the continuum limit can be considered to be reliable.

\end{document}